\documentclass
[twocolumn,superscriptaddress,secnumarabic,amssymb,amsmath,nobibnotes,aps,prd,showpacs,nofootinbib]{revtex4}%
\usepackage{graphicx}
\usepackage{epsf}
\usepackage{bm}
\usepackage{amsmath}
\usepackage{amsfonts}
\usepackage{amssymb}
\usepackage{epstopdf}
\usepackage{natbib}
\usepackage{rotating}
\usepackage{pdflscape}
\usepackage{adjustbox}
\usepackage{color}%
\providecommand{\U}[1]{\protect\rule{.1in}{.1in}}
\newcommand{\be}{\begin{equation}}
\newcommand{\ee}{\end{equation}}

\newcommand{\mincir}{\raise
-3.truept\hbox{\rlap{\hbox{$\sim$}}\raise4.truept\hbox{$<$}\ }}
\newcommand{\magcir}{\raise
-3.truept\hbox{\rlap{\hbox{$\sim$}}\raise4.truept\hbox{$>$}\ }}

\begin{document}
\title{Latest astronomical constraints on some nonlinear parametric dark energy models}
\author{Weiqiang Yang}
\email{d11102004@163.com}
\affiliation{Department of Physics, Liaoning Normal University, Dalian, 116029, P. R. China}
\author{Supriya Pan}
\email{span@research.jdvu.ac.in}
\affiliation{Department of Physical Sciences, Indian Institute of Science Education and
Research, Kolkata, Mohanpur$-$741246, West Bengal, India}
\affiliation{Department of Mathematics, Raiganj Surendranath Mahavidyalaya, Sudarshanpur,
Raiganj, West Bengal 733134, India}
\author{Andronikos Paliathanasis}
\email{anpaliat@phys.uoa.gr}
\affiliation{Instituto de Ciencias F\'{\i}sicas y Matem\'{a}ticas, Universidad Austral de Chile, 5090000 Valdivia, Chile}
\affiliation{Institute of Systems Science, Durban University of Technology, PO Box 1334,
Durban 4000, Republic of South Africa}


\begin{abstract}
We consider nonlinear redshift-dependent equation of state parameters as dark
energy models in a spatially flat Friedmann-Lema\^{\i}tre-Robertson-Walker
universe. To depict the expansion history of the universe in such cosmological
scenarios, we take into account the large scale behaviour of such parametric
models and fit them using a set of latest observational data with distinct
origin that includes cosmic microwave background radiation, Supernove Type Ia,
baryon acoustic oscillations, redshift space distortion, weak gravitational
lensing, Hubble parameter measurements from cosmic chronometers and finally
the local Hubble constant from Hubble space telescope. The fitting technique
avails the publicly available code Cosmological Monte Carlo (CosmoMC), to
extract the cosmological information out of these parametric dark energy models. 
From our analysis it follows that those models could describe the late time
accelerating phase of the universe, while they are distinguished from the
$\Lambda-$cosmology. 

\end{abstract}
\maketitle



\section{Introduction}

Astronomical observations from various independent sources consistently
suggest that our Universe is expanding in an accelerating manner
\cite{Riess:1998cb,Perlmutter:1998np,Tegmark:2003uf,
Eisenstein:2005su, Bennett:2012zja, Ade:2015xua}. Various approaches have been
proposed in the literature. Those apporaches can be classified into two big
categories. The dark energy models in which an exotic fluid is introduced in
Einstein grvaity, and the modified theories of gravity in which
geometrodynamical quantities are introduced in the gravitational action to
explain the accelerating phase of the universe. Of course these two categories
include in common the cosmological constant. Because the latter can be seen
either as a perfect fluid with negative equation of state parameter or as a
modification of the Einstein-Hilbert Action.

Here we are interested on the parametric dark energy models. In
particular, we consider the existence of an exotic dark energy fluid with
time-varying equation of state parameter in order to perform the fittings of
the cosmological observations. The theoretical origin of those parametric
equation of state parameters can be from the dark energy models
\cite{Copeland:2006wr,Ratra:1987rm,Carroll:1998zi,
UrenaLopez:2000aj,Kamenshchik:2001cp,deHaro:2016hpl, Sharov:2015ifa,Yang:2017amu,
Yang:2017zjs}, or from the modified theories of
gravity \cite{Nojiri:2017ncd, DeFelice:2010aj, Li:2007xn, paliagrg, sotiriou,
Paliathanasis:2016vsw, Cai:2015emx, Nunes:2016drj, Nunes:2016qyp, Lu:2016hsd} (see also
the references therein).

In the last decades different parametrizations for the dark energy equation of
state have been proposed in the literature 
which include some well known parametrizations, namely, linear parametrization \cite{lin1, lin2, lin3}, Chevallier-Polarski-Linder (known as CPL) parametrization \cite{cpl1,cpl2},
logarithmic parametrization \cite{log}, Jassal-Bagla-Padmanabhan (known as JBP)
parametrization \cite{Jassal:2005qc}, Barboza-Alcaniz 
parametrization \cite{Barboza:2008rh}, and many more,
see also \cite{periv1, periv2, Linder:2005ne,Zhang:2015lda,Nesseris:2005ur,Pantazis:2016nky} and the references therein.

In this work we consider that the fluids of the dark energy sector are
\ non-interacting while for the underlying geometry we consider  the
spatially flat Friedmann-Lema\^{\i}tre-Robertson-Walker (FLRW) spacetime.
Moreover, we assume that the total energy density of the universe is shared by
baryons, radiation, dark matter and dark energy. The dark matter is considered
to be pressureless while the dark energy fluid is barotropic and represented
by some parametric models. To check the robustness of those parametric dark
energy models we consider their large scale behaviour and fit them using the
latest observational data from different astronomical probes, namely, cosmic
microwave background radiation \cite{ref:Planck2015-1, ref:Planck2015-2},
Supernove Type Ia sample \cite{Betoule:2014frx}, baryon acoustic oscillation
distance measurements
\cite{ref:BAO1-Beutler2011,ref:BAO2-Ross2015,ref:BAO3-Gil-Marn2015}, redshift
space distortion \cite{ref:BAO3-Gil-Marn2015}, weak gravitational lensing
\cite{Heymans:2013fya,Asgari:2016xuw} and finally the history of the Hubble
parameter that combines the latest complilation of the cosmic chronometers
\cite{Moresco:2016mzx} and the local Hubble constant \cite{Riess:2016jrr}.
From our analysis we see that the parametric dark energy models could be an
effective way to study the late time accelerating phase of the universe. The
observational constraints on the models suggest that the current value of the
dark energy equation of state is close to that of the 
cosmological constant but \ the
models are distinguished between them and from the cosmological constant 
as well.

We discuss the background cosmology and introduce the parameteric dark energy
models in section \ref{sec-setup}. The perturbation equations for the dark
energy models have been written in section \ref{sec-perturbations}. The
observational data used in our analysis have been summarized in section
\ref{sec-data} and consequenlty the section \ref{sec-results} contains the
results of our statistical analysis. Finally, we close our work in section
\ref{discuss} with a brief outline of the whole work.

\section{Parametric dark energy models}

\label{sec-setup}

The cosmological scenario of our consideration is that of a four dimensional
spacetime which is isotropic, homogeneous and spatially flat. In particular
the line element, i.e. the metric, is invariant under the action of the
Euclidean group $E^{3}$. The corresponding line element of the FLRW
spacetime is,%
\[
ds^{2}=-N^{2}\left(  t\right)  dt^{2}+a(t)^{2}\left(  dr^{2}+r^{2}\left(
d\theta^{2}+\sin^{2}\theta d\phi^{2}\right)  \right),
\]
where $a\left(  t\right)  $ is the scalar factor of the universe and $N\left(
t\right)  $ is the lapse function which without any loss of generality we can
consider that $N\left(  t\right)  =1$. Furthremore, we consider the comoving
observer $u^{a}=\frac{1}{N\left(  t\right)  }\delta_{t}^{a}$, such that
$u^{a}u_{a}=-1$.

For such spacetime the Einstein's field equation are calculated as%

\begin{equation}
H^{2}=\frac{8\pi G}{3}(\rho_{r}+\rho_{b}+\rho_{c}+\rho_{x}), \label{f1}%
\end{equation}

\begin{equation}
2\dot{H}+3H^{2}=-8\pi G(p_{r}+p_{b}+p_{c}+p_{x}), \label{f2}%
\end{equation}
where $H=\dot{a}/a$, is the Hubble rate of this universe; $(\rho_{i},p_{i})$
stand respectively, for the energy density and the pressure of the $i$-th
fluid where $i$ runs over  radiation ($r$), baryons ($b$), dark matter $(c$)
and a dark energy fluid ($x$).

Furthermore, we consider that all the fluid components are minimally coupled
to gravity and there is no interaction between the fluids. Hence, the
Bianchi identity leads to $\nabla_{\nu}T^{\mu\nu}=0$, in which the radiation and baryons follow the evolution laws $\rho_{r}\propto\left( a/a_{0} \right) ^{-4}$, $\rho_{b} \propto\left( a/a_{0} \right) ^{-3}$, respectively, while for the other two components, we have%

\[
\rho_{c,x}=\rho_{c,x}^{0}\,\left(  \frac{a}{a_{0}}\right)  ^{-3}\,\exp\left(
-3\int_{a_{0}}^{a}\frac{w_{c,x}\left(  \sigma\right)  }{\sigma}\,d\sigma
\right)~.
\]

This is a general evolution equation for both dark matter and dark energy with
equations of state $w_{c}$, $w_{x}$, respectively where $\rho_{c,x}^{0}$ is a
shorthand notation that describes both $\rho_{c}^{0}$ and $\rho_{x}^{0}$ as
the present values of the corresponding energy densities, $a_{0}$ is the
present value of the scale factor which is related to the cosmological
redshift $z$, as $a_{0}a^{-1}=1+z$. For constant $w_{c}$, the above equation
becomes $\rho_{c}=\rho_{c}^{0}\left(  a/a_{0}\right)  ^{-3(1+w_{c})}$. In this
work, we shall consider pressureless dark matter, that is $w_{c}=0$, but the 
equation of state for DE is
dynamical (time-varying, i.e. $w_{x}=w_{x}(a\left(  t\right)  )$) while in
particular we consider four nonlinear functions $w_{x}\left(  a\right)  $,
where three of which have been proposed before in the litterature and 
we propose a new parameterization in this work.

The scope of our analysis is to study the viability of those models with the
use of the recent cosmological data, and also to determine the best parameters
of the models. Let us remark that throughout the work we set $a_{0}=1$.

\subsection{Model A}

As the first model we consider the Barboza-Alcaniz equation of state parameter
\cite{Barboza:2008rh} which has two free parameters%

\begin{equation}
w_{x}(z)=w_{0}+w_{a}\frac{z(1+z)}{1+z^{2}}. \label{BA}%
\end{equation}
Parameter $w_{0}~$gives the value of $w_{x}~$at present, that is $w_{0}%
=w_{x}\left(  z\rightarrow0\right)  $, while for large values of $z$,
$w_{x}\left(  z\right)  $ becomes constant. Moreover, ~the parameter $w_{a}$
denotes the rate of change of the equation of state parameter today, that is,
$w_{a}=\frac{dw_{x}}{dz}|_{z\rightarrow0}$. One can easily realize that this
parametrization is divergence free.

\begin{figure}
\includegraphics[width=8.0cm,height=7.5cm]{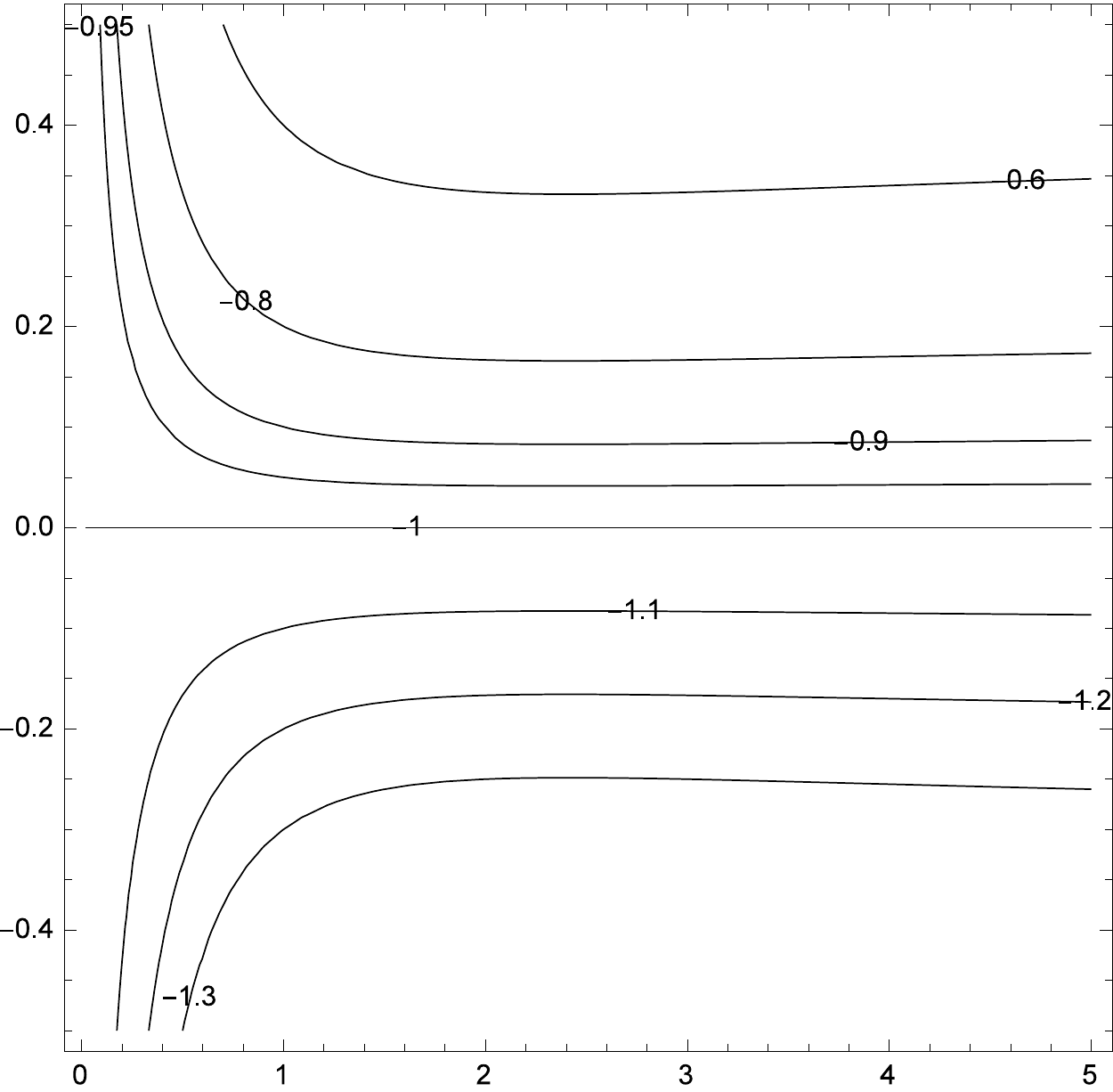} \caption{Contour plots
for the parametric model $A$ (Barboza-Alcaniz parametrization) of (\ref{BA}) is
presented for $w_{0} = -1$. Horizontal axis presents the redshift history while
the vertical axis takes the values of the constant $w_{a}$. }%
\label{modelafig}%
\end{figure}

In Fig. \ref{modelafig} the evolution of the the parametric model is presented
in terms of the redshift and the second parameter $w_{a}.$

\subsection{Model B}

A nonlinear model for the equation of state parmater of dark energy was
proposed in \cite{Zhang:2015lda}. The functional form of the
model as presented in \cite{Zhang:2015lda} is%
\[
w_{x}(z)=-1+\frac{1}{3}\left(  \frac{\beta(1+z)}{\alpha+\beta(1+z)}\right)
\]
where parameters $\alpha~$and $\beta\ $are positive real numbers. However,
although the model contains two free parameters but it is easy to see that
there is only one real free parameter.

In particular, we consider a natural generalization of the model and we
consider the following functional form%
\begin{equation}
w_{x}(z)=w_{1}+\frac{1}{3}\left(  \frac{(1+z)}{w_{2}+(1+z)}\right)  ,
\label{Zhang1}%
\end{equation}
where $w_{1}$, $w_{2}~(\neq 0,~-1)$, are the free parameters\footnote{It must
be mentioned that $w_{2}$ cannot take $0,~-1$ because, for $w_{2}=0$,
$w_{x}(z)$ becomes constant while for $w_{2}=-1$, the model diverges at $z=0$,
which is unphysical.}. Now considering the present value of the dark energy
equation of state, $w_{0}$, for the parametrization (\ref{Zhang1}) one has
$w_{0}=w_{1}+\frac{1}{3\left(  w_{2}+1\right)  }$. However, we observe that
for large redshifts the equation of state parameter for the model becomes
constant in a similar way with that of the first model (\ref{BA}).

It is easy to see that using the current value of the dark energy equation of
state $w_{0}$, the parametrization (\ref{Zhang1})  can  be
recast as%

\begin{equation}
w_{x}(z)=\left(  w_{0}-\frac{1}{3\left(  w_{2}+1\right)  }\right)  +\frac
{1}{3}\left(  \frac{(1+z)}{w_{2}+(1+z)}\right)  , \label{Zhang1.1}%
\end{equation}

\begin{figure}
\includegraphics[width=8.0cm,height=7.5cm]{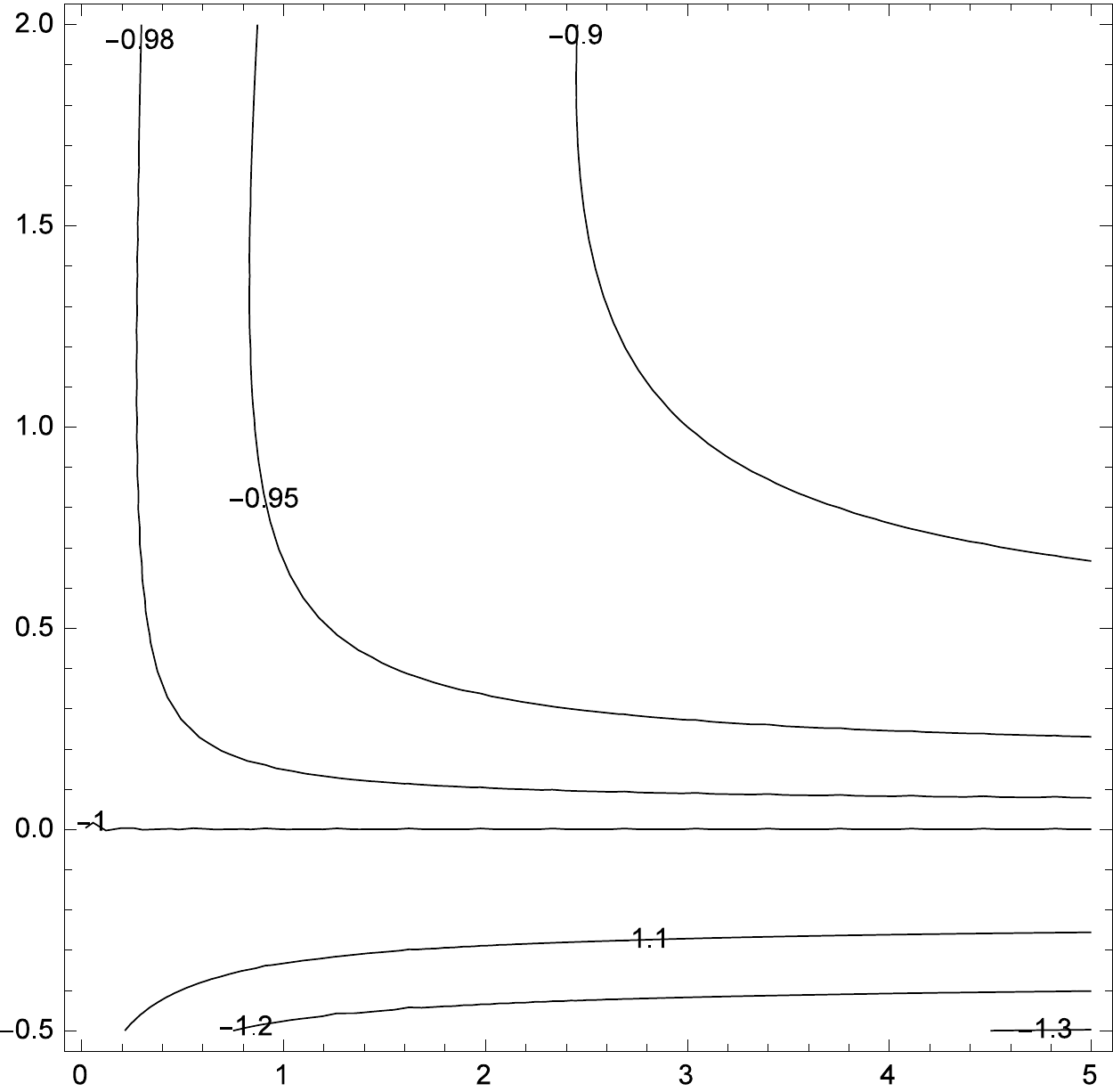} \caption{Contour plots
for the parametric model $B$ (Zhang et al. parametrization) of (\ref{Zhang1.1})
is presented for the free parameter $w_{0}=-1$. Horizontal axis presents the
redshift history while the vertical axis provides the second parameter $w_{2}%
$. }%
\label{modelbfig}%
\end{figure}

We note that the model (\ref{Zhang1.1}) may diverge at finite time given by
$z=-1-w_{2}$. However, we make it clear that for $w_{2}>0$ the model does not
show any diverging nature in past or future. The model sustains its
diverging character only for the negative values of $w_{2}$. Precisely, one
can find that (\ref{Zhang1.1}) could diverge at past or future, respectively
for $w_{2}<-1$ or $-1<w_{2}<0$.~Moreover, in Fig. \ref{modelbfig} we present
the evolution of the parametric equation of state parmeter in terms of the
cosmic history and the free parameter $w_{2}$.

\subsection{Model C}

We introduce another model that was originally introduced in
\cite{Nesseris:2005ur}. We note that except the nonlinear
character, the equation of state parameter depends on the energy density of
the matter sector (cold dark matter plus baryons), that is,%

\begin{equation}
w_{x}(z)=\frac{a_{1}+3(\Omega_{m0}-1)-2a_{1}z-a_{2}(z^{2}+2z-2)}%
{3(1-\Omega_{m0}+a_{1}z+2a_{2}z+a_{2}z^{2})}, \label{nnl}%
\end{equation}
where $a_{1}$, $a_{2}$ are free constants and $\Omega_{m0}$, is the current
density parameter for matter sector.

One can observe that both for $a_{1}=a_{2}=0$, $w_{x} (z)=-1$. Thus, in order for
(\ref{nnl}) to represent a dynamical equation of state for the
dark energy model, one of the free
parameters, namely $a_{1}$, $a_{2}$ must be 
nonzero. However, since (\ref{nnl}) is a
rational function, so one may realize that the model could exhibit past or
future singularities depending on the free parameters. Let us first consider
the simplest cases when only one of $a_{1}$ and $a_{2}$ is zero. For instance,
if we consider that $a_{2}=0$, but $a_{1}\neq0$, we see that the model
(\ref{nnl}) gives a singularity at $z=-\left(  \frac{1-\Omega_{0}}{a_{1}%
}\right)  $. Now, this sigularity could be at past or present depending on the
nature of $a_{1}$. For $a_{1}<0$, the singularity appears at $z>0$, that means
a past singularity while for $a_{1}>0$, the singularity arises at $z<0$
menaing a future singularity. We consider the second possibility when
$a_{1}=0$, but $a_{2}\neq0$. Here two distinct possibilities arise $-$ when
$a_{2}^{2}-a_{2}(1-\Omega_{m0})<0$, we do not have any singularity but for
$a_{2}^{2}-a_{2}(1-\Omega_{m0})>0$ we have singularities. Now for this second
case, if $a_{2}<0$, then we always have a singularity given by $z=-1-\frac
{1}{a_{2}}\sqrt{a_{2}^{2}-a_{2}(1-\Omega_{m0})}$. This singularity can be
classified as past or future singularity whether $z>0$, or $-1<z<0$,
respectively. On the other hand, for $a_{2}>0$, singularity exists for
$a_{2}>(1-\Omega_{m0})$ and the singularity occurs at $z=-1+\frac{1}{a_{2}%
}\sqrt{a_{2}^{2}-a_{2}(1-\Omega_{m0})}$. One can similarly classify this
singularity as described just above. Now we consider the general case when
$a_{1}\neq0$ and $a_{2}\neq0$. We notice that the parametrization could have
real singularities if $(2a_{2}+a_{1})^{2}-4a_{2}(1-\Omega_{m0})>0$. However,
for $(2a_{2}+a_{1})^{2}-4a_{2}(1-\Omega_{m0})<0$, one does not encouter with
any real singularities in the above parametrization.

\begin{figure}
\includegraphics[width=8.0cm,height=7.5cm]{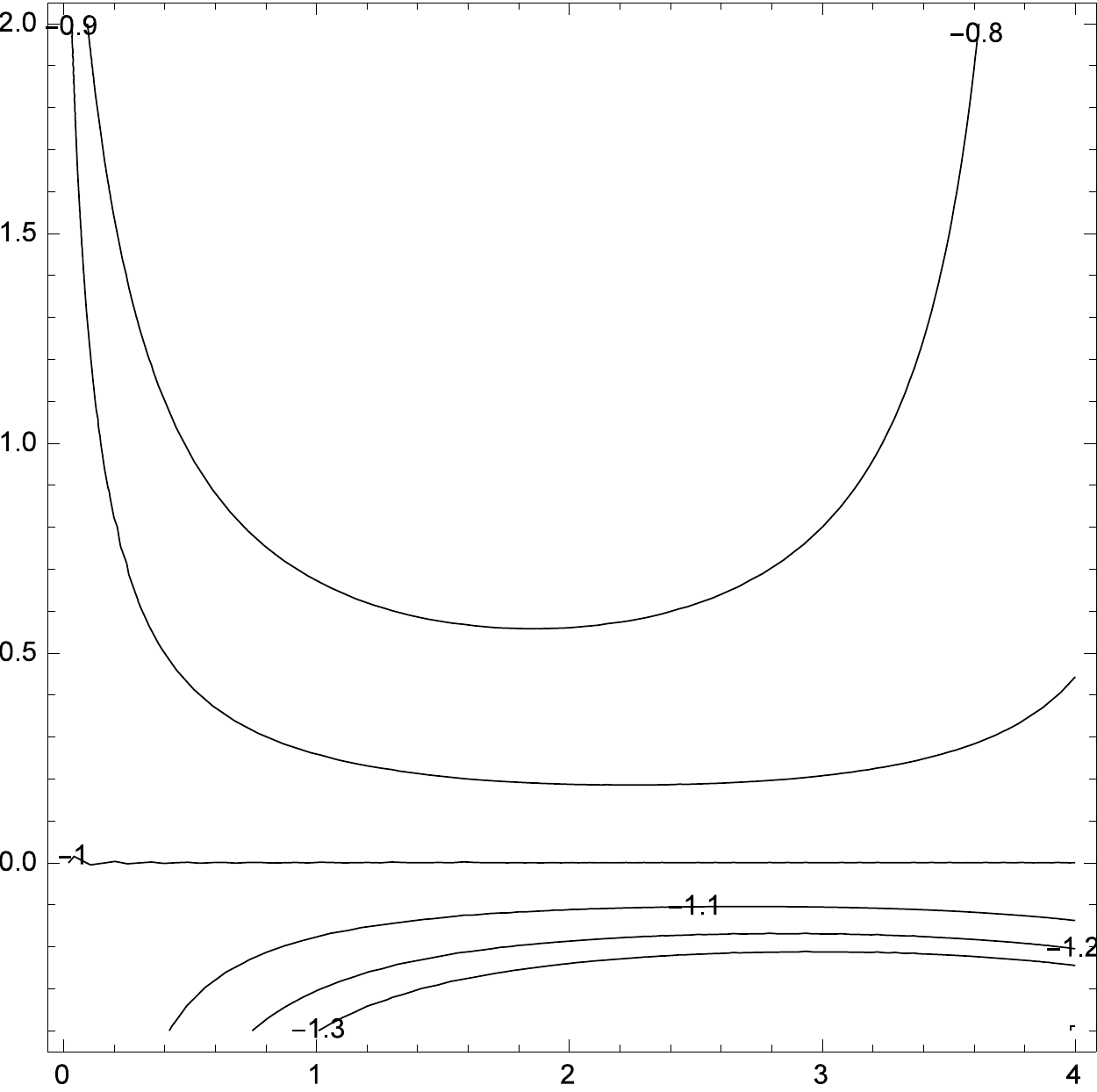} \caption{Contour plots
for the parametric model $C$ (Nesseris and Perivolaropoulos parametrization) of
(\ref{nnl}) is presented for the free parameter $\Omega_{m0}=0.30$ and
$w_{x}\left(  z\rightarrow0\right)  =-1$. Horizontal axis presents the
redshift history while the vertical axis provides the free parameter $a_{1}$.
}%
\label{modelfigc}%
\end{figure}

The evolution of this equation of state parameter is displayed in Fig.
\ref{modelfigc} for $\Omega_{m0}=0.30$ and the constraint $w_{x}\left(
a\rightarrow1\right)  =-1$.

\subsection{Model D}

We introduce the final model in this series having the following form%

\begin{equation}
w_{x}(z)=w_{0}+\frac{w_{b}}{1+z}\ln(1+z) \label{model-d}%
\end{equation}
where $w_{0}$ is the current value of $w_{x}(z)$ and $w_{b}$ is any free
parameter. We note that for $z\rightarrow\infty$, $w_{x} (z)\rightarrow w_{0}$.
One can see that at low redshifts, the model (\ref{model-d}) recovers the CPL
parametrization $w_{x}(z)=w_{0}-w_{b}\left(  \frac{z}{1+z}\right)  $. That
means at very low redshifts, the evolution of (\ref{model-d}) and the evolution
of the CPL parametrization are equivalent.

\begin{figure}
\includegraphics[width=8.0cm,height=7.5cm]{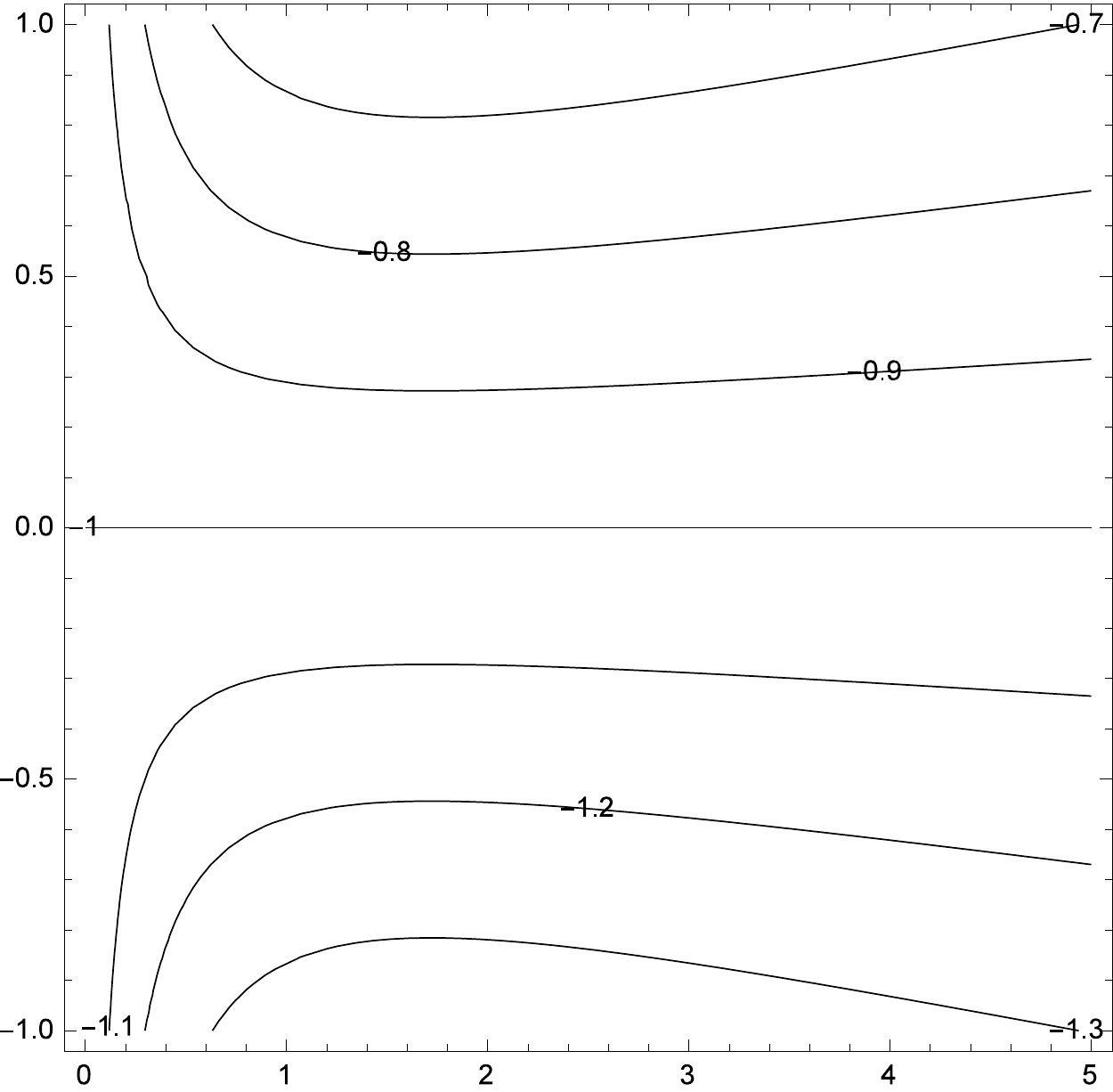}\caption{Contour plots for
the parametric model D (The new parametrization) of (\ref{model-d}) is
presented for $w_{0}=-1$. Horizontal axis presents the redshift history while
the vertical axis takes the values of the constant $w_{b}$. }%
\label{modeldfig}%
\end{figure}

In Fig. \ref{modeldfig} we present the evolution of the parametric equation of
state parmeter in terms of the cosmic history and the free parameter $w_{b}$.

\section{Dynamical Dark-Energy at perturbative level}

\label{sec-perturbations}

The most general scalar mode perturbation is characterized by the following
metric \cite{Mukhanov, Ma:1995ey, Malik}%

\begin{align}
ds^{2} = -(1+ 2 \phi) dt^{2} + 2 a \partial_{i} B dt dx +\nonumber\\
a^{2} [(1-2 \psi) \delta_{ij} + 2 \partial_{i} \partial_{j} E] dx^{i} dx^{j}.
\end{align}
In this work we follow some standard formalism as described in Ref.
\cite{Ma:1995ey}, and adopt the synchronous gauge, that means we consider
$\phi= B = 0$, $\psi= \eta$, and $k^{2} E = - h/2 - 3 \eta$. Using the above
metric in consideration of the synchronous gauge, the conservation equations
for the $i$-th component of the fluid are found to be
\begin{align}
\delta^{\prime}_{i}  &  = - (1+ w_{i})\, \left(  \theta_{i}+ \frac{h^{\prime}%
}{2}\right)  - 3 \mathcal{H} \left(  \frac{\delta P_{i}}{\delta\rho_{i}} -
w_{i} \right)  \delta_{i}\,,\\
\theta^{\prime}_{i}  &  = - \mathcal{H} (1- 3w_{i})\theta_{i}- \frac
{w^{\prime}_{i}}{1+w_{i}} \theta_{i} + \frac{\delta P_{i}/\delta\rho_{i}%
}{1+w_{i}}\, k^{2}\, \delta_{i}- k^{2} \, \sigma.
\end{align}
Here the prime is the derivative with respect to the conformal time; $\mathcal{H}
= a^{\prime}/a$, is the conformal Hubble parameter; and the remaining
quantities $\sigma$, $\delta_{i}$, $\theta_{i}$, are respectively denote the
shear, density perturbation and the velocity pertubation. The DE perturbation
equations follow
\begin{align}
\delta^{\prime}_{x}  &  = - (1+ w_{x})\, \left(  \theta_{x}+ \frac{h^{\prime}%
}{2}\right)  - 3\mathcal{H}w^{\prime}_{x}\frac{\theta_{x}}{k^{2}}\nonumber\\
&  - 3 \mathcal{H} \left(  c^{2}_{s} - w_{x} \right)  \left[  \delta
_{x}+3\mathcal{H}(1+w_{x})\frac{\theta_{x}}{k^{2}} \right]  \,,\\
\theta^{\prime}_{x}  &  = - \mathcal{H} (1- 3c^{2}_{s})\theta_{x} +
\frac{c^{2}_{s}}{1+w_{x}}\, k^{2}\, \delta_{x}%
\end{align}
where without any loss of generality and simplicity we assume $\sigma=0$;
$c^{2}_{s}$, is the physical sound speed in the rest frame which we set to be
unity to avoid any kind of unphysical behaviour. The perturbation equations
for the other components, namely, the baryons, radiation, and pressureless
dark matter follow the standard equations as described in \cite{Mukhanov,
Ma:1995ey, Malik}.

\begingroup

\begin{table}
\begin{tabular}
[c]{cccccc}\hline\hline
Parameters & Priors &Mean with errors & Best fit\\\hline
$\Omega_{c} h^{2}$ & $[0.01, 0.99]$ &$0.1186_{- 0.0013- 0.0025}^{+ 0.0013+
0.0024}$ & $0.1195$\\
$\Omega_{b} h^{2}$ & $[0.005, 0.1]$ &$0.0223_{- 0.0002- 0.0003}^{+ 0.0001+
0.0003}$ & $0.0223$\\
$100\theta_{MC}$ & $[0.5, 10]$ &$1.0406_{- 0.0003- 0.0006}^{+ 0.0003+
0.0006}$ & $1.0406$\\
$\tau$ &  $[0.01, 0.8]$ &$0.0636_{- 0.0173- 0.0338}^{+ 0.0170+ 0.0341}$ &
$0.0502$\\
$n_{s}$ & $[0.5, 1.5]$ &$0.9753_{- 0.0044- 0.0085}^{+ 0.0044+ 0.0090}$ &
$0.9718$\\
$\mathrm{{ln}}(10^{10} A_{s})$ & $[2.4, 4]$ &$3.0676_{- 0.0335- 0.0640}^{+
0.0326+ 0.0661}$ & $3.0468$\\\hline
$w_{0}$ & $[-2, 0]$ &$-0.9589_{- 0.0910- 0.1796}^{+ 0.0871+ 0.1797}$
& $-0.9112$\\
$w_{a}$ & $[-1, 1]$ &$-0.1536_{- 0.1623- 0.3740}^{+ 0.2109+ 0.3705}$
& $-0.2828$\\\hline
$\Omega_{m0}$ & $-$ &$0.3031_{- 0.0091- 0.0158}^{+ 0.0079+
0.0168}$ & $0.3051$\\
$\sigma_{8}$ & $-$ & $0.8197_{- 0.0132- 0.0271}^{+ 0.0144+
0.0268}$ & $0.8177$\\
$H_{0}$ & $-$ & $68.3496_{- 0.8664- 1.6475}^{+ 0.8667 + 1.6581}$
& $68.3179$\\\hline
$\chi^{2}_{min}$ & $-$ &  $-$ & 13720.556\\\hline\hline
\end{tabular}
\caption{\textit{The table summarizes the observational constraints of the
free and derived parameters of the Model $A$ (Barboza-Alcaniz model) for the
combined observational data Planck TT, TE, EE $+$ lowTEB $+$ JLA $+$ BAO $+$
RSD $+$ WL $+$ CC $+$ $H_{0}$. Here, $\Omega_{m0}=\Omega_{c0}+\Omega_{b0}$. }}%
\label{tab:BA}%
\end{table}

\endgroup

\begin{figure}
\includegraphics[width=8cm,height=7.5cm]{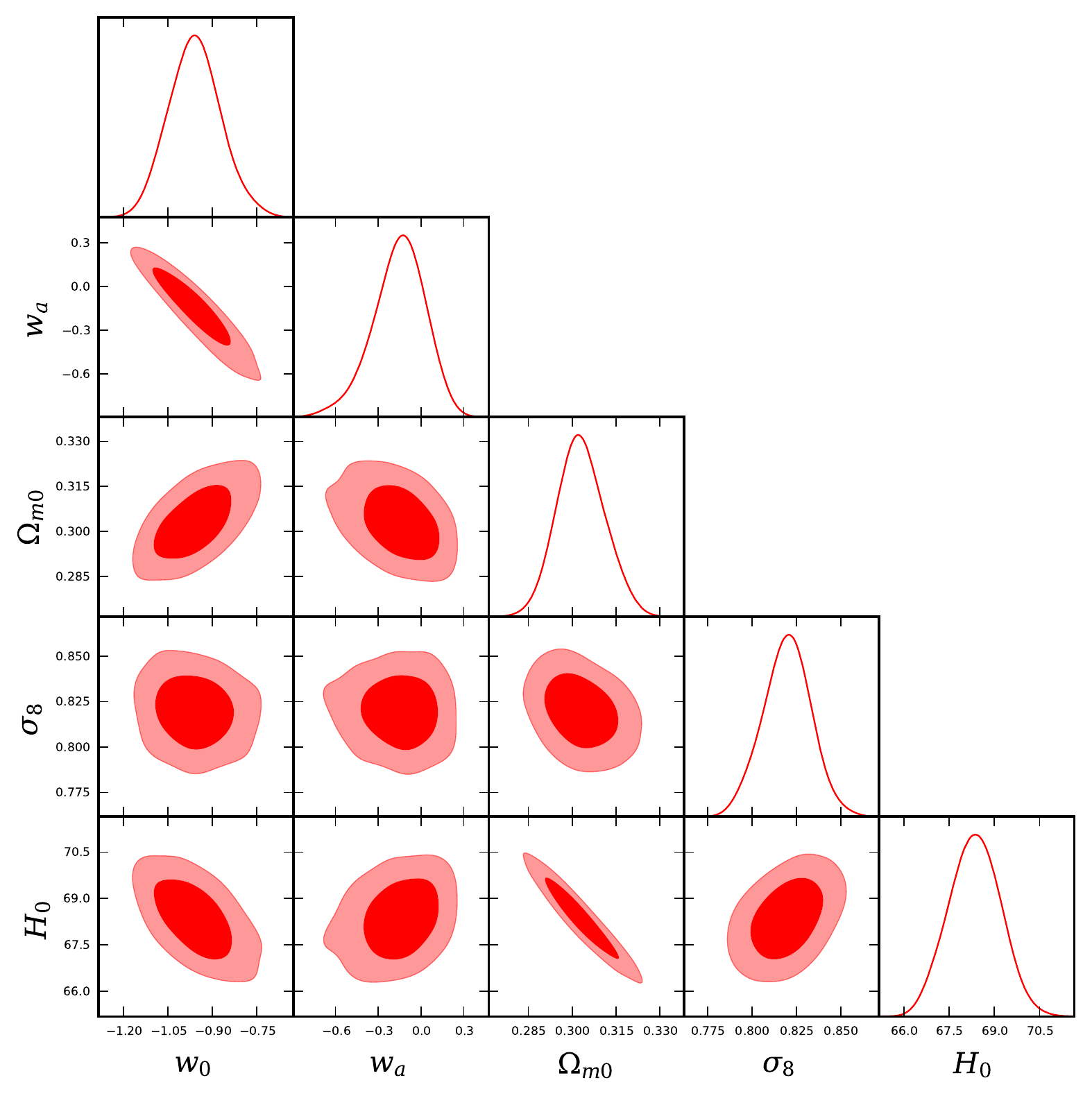}
\caption{\textit{The figure displays the 68.3\% and 95.4\% confidence-level
contour plots for different combinations of the model parameters of the
Barboza-Alcaniz parametrization (\ref{BA}) using the combined analysis Planck TT, TE, EE
$+$ lowTEB $+$ JLA $+$ BAO $+$ RSD $+$ WL $+$ CC $+$ $H_{0}$. }}%
\label{fig:BA}%
\end{figure}

\begingroup
\begin{table}
\begin{tabular}
[c]{ccccc}\hline\hline
Parameters & Priors & Mean with errors & Best fit\\\hline
$\Omega_{c} h^{2}$ & $[0.01, 0.99]$ & $0.1182_{- 0.0013- 0.0023}^{+ 0.0012+
0.0023}$ & $0.1183$\\
$\Omega_{b} h^{2}$ & $[0.005, 0.1]$ & $0.0223_{- 0.0002- 0.0003}^{+ 0.0001+
0.0003}$ & $0.0223$\\
$100\theta_{MC}$ & $[0.5, 10]$ &$1.0406_{- 0.0003- 0.0006}^{+ 0.0003+
0.0006}$ & $1.0408$\\
$\tau$ & $[0.01, 0.8]$ &$0.0662_{- 0.0182- 0.0324}^{+ 0.0168+ 0.0344}$ &
$0.0688$\\
$n_{s}$ &  $[0.5, 1.5]$ &$0.9762_{- 0.0043- 0.0075}^{+ 0.0039+ 0.0081}$ &
$0.9787$\\
$\mathrm{{ln}}(10^{10} A_{s})$ & $[2.4, 4]$ &$3.0721_{- 0.0349- 0.0636}^{+
0.0328+ 0.0668}$ & $3.0799$\\\hline
$w_{0}$ & $[0, 2]$ & $-1.0449_{- 0.0322- 0.0717}^{+ 0.0361+ 0.0677}$
& $-1.0416$\\
$w_{2}$ & $[-3, 3]$ &$0.2254_{- 0.2254- 0.2254}^{+ 0.0509+ 0.3364}$ &
$0.0220$\\\hline
$\Omega_{m0}$ & $-$ &$0.2991_{- 0.0079- 0.0142}^{+ 0.0072+
0.0147}$ & $0.2971$\\
$\sigma_{8}$ & $-$ &$0.8198_{- 0.0139- 0.0276}^{+ 0.0147+
0.0266}$ & $0.8269$\\
$H_{0}$ & $-$ &$68.7052_{- 0.8120- 1.6566}^{+ 0.8192+ 1.6440}$
& $68.9673$\\\hline
$\chi^{2}_{min}$ & $-$ & $-$ & 13722.266\\\hline\hline
\end{tabular}
\caption{\textit{The table summarizes the observational constraints of the
free and derived parameters of Model $B$ 
(\ref{Zhang1.1}) (Zhang et al. parametrization)  for the
combined observational data Planck TT, TE, EE $+$ lowTEB $+$ JLA $+$ BAO $+$
RSD $+$ WL $+$ CC $+$ $H_{0}$. We note that, $\Omega_{m0}=\Omega_{c0}%
+\Omega_{b0}$.}}%
\label{tab:ZhangI}%
\end{table}
\endgroup

\begin{figure}
\includegraphics[width=8cm,height=7.5cm]{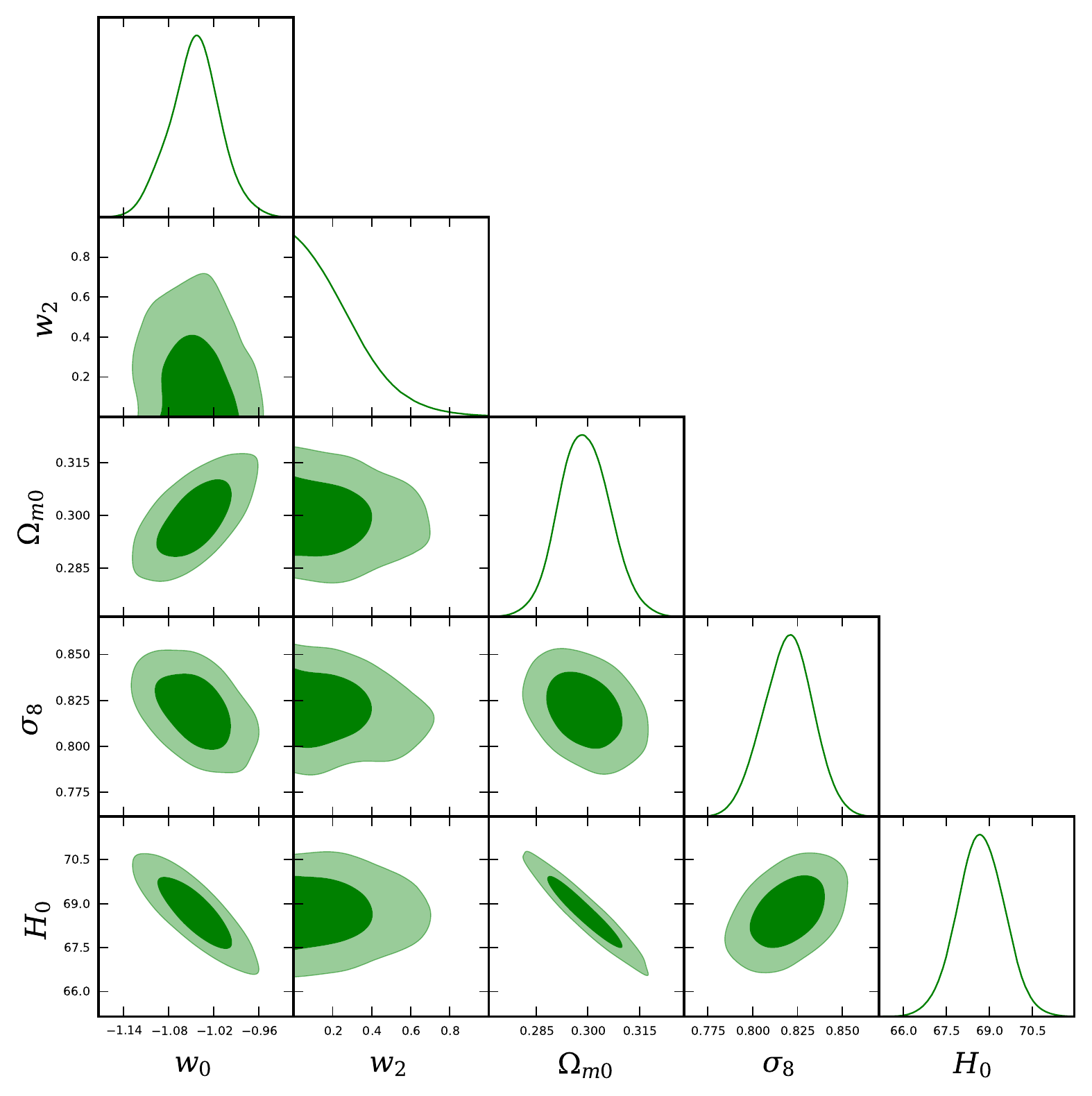}
\caption{\textit{The figure displays the 68.3\% and 95.4\% confidence-level
contour plots for different combinations of the model parameters of the Zhang
et al. parametrization (\ref{Zhang1.1}) using the combined analysis Planck TT, TE, EE $+$ lowTEB $+$ JLA $+$ BAO $+$ RSD $+$ WL $+$ CC $+$ $H_{0}$. }}%
\label{fig:ZhangI}%
\end{figure}

\begingroup
\begin{table}
\begin{tabular}
[c]{ccccc}\hline\hline
Parameters & Priors &Mean with errors & Best fit\\\hline
$\Omega_{c} h^{2}$ & $[0.01, 0.99]$ &$0.1182_{- 0.0013- 0.0024}^{+ 0.0012+
0.0025}$ & $0.1186$\\
$\Omega_{b} h^{2}$ & $[0.005, 0.1]$ & $0.0223_{- 0.0002- 0.0003}^{+ 0.0002+
0.0003}$ & $0.0222$\\
$100\theta_{MC}$ & $[0.5, 10]$ &$1.0406_{- 0.0003- 0.0006}^{+ 0.0003+
0.0006}$ & $1.0405$\\
$\tau$ & $[0.01, 0.8]$ &$0.0687_{- 0.0165- 0.0344}^{+ 0.0166+ 0.0328}$ &
$0.0604$\\
$n_{s}$ & [0.5, 1.5] &$0.9761_{- 0.0042- 0.0084}^{+ 0.0045+ 0.0083}$ &
$0.9768$\\
$\mathrm{{ln}}(10^{10} A_{s})$ & $[2.4, 4]$ &$3.0759_{- 0.0325- 0.0640}^{+
0.0324+ 0.0627}$ & $3.0658$\\\hline
$a_{1}$ & $[-1, 0]$ &$-0.1379_{- 0.0548- 0.1613}^{+ 0.1107+ 0.1379}$
& $-0.1113$\\
$a_{2}$ & $[0, 1]$ &$0.0244_{- 0.0244- 0.0244}^{+ 0.0058+ 0.0366}$ &
$0.0138$\\\hline
$\Omega_{m0}$ & $-$ &$0.2984_{- 0.0073- 0.0137}^{+ 0.0074+
0.0144}$ & $0.2990$\\
$\sigma_{8}$ & $-$ & $0.8195_{- 0.0130- 0.0268}^{+ 0.0135+
0.0252}$ & $0.8222$\\
$H_{0}$ & $-$ & $68.7759_{- 0.8318- 1.5894}^{+ 0.7928+ 1.5598}$
& $68.7964$\\\hline
$\chi^{2}_{min}$ & $-$ & $-$ & 13723.394\\\hline\hline
\end{tabular}
\caption{\textit{The table summarizes the observational constraints of the
free and derived parameters of the Model $C$ (Nesseris-Perivolaropoulos
parametrization) for the combined observational data Planck TT, TE, EE $+$
lowTEB $+$ JLA $+$ BAO $+$ RSD $+$ WL $+$ CC $+$ $H_{0}$. We note that,
$\Omega_{m0}=\Omega_{c0}+\Omega_{b0}$.}}%
\label{tab:NP}%
\end{table}

\endgroup

\begin{figure}
\includegraphics[width=8cm,height=7.5cm]{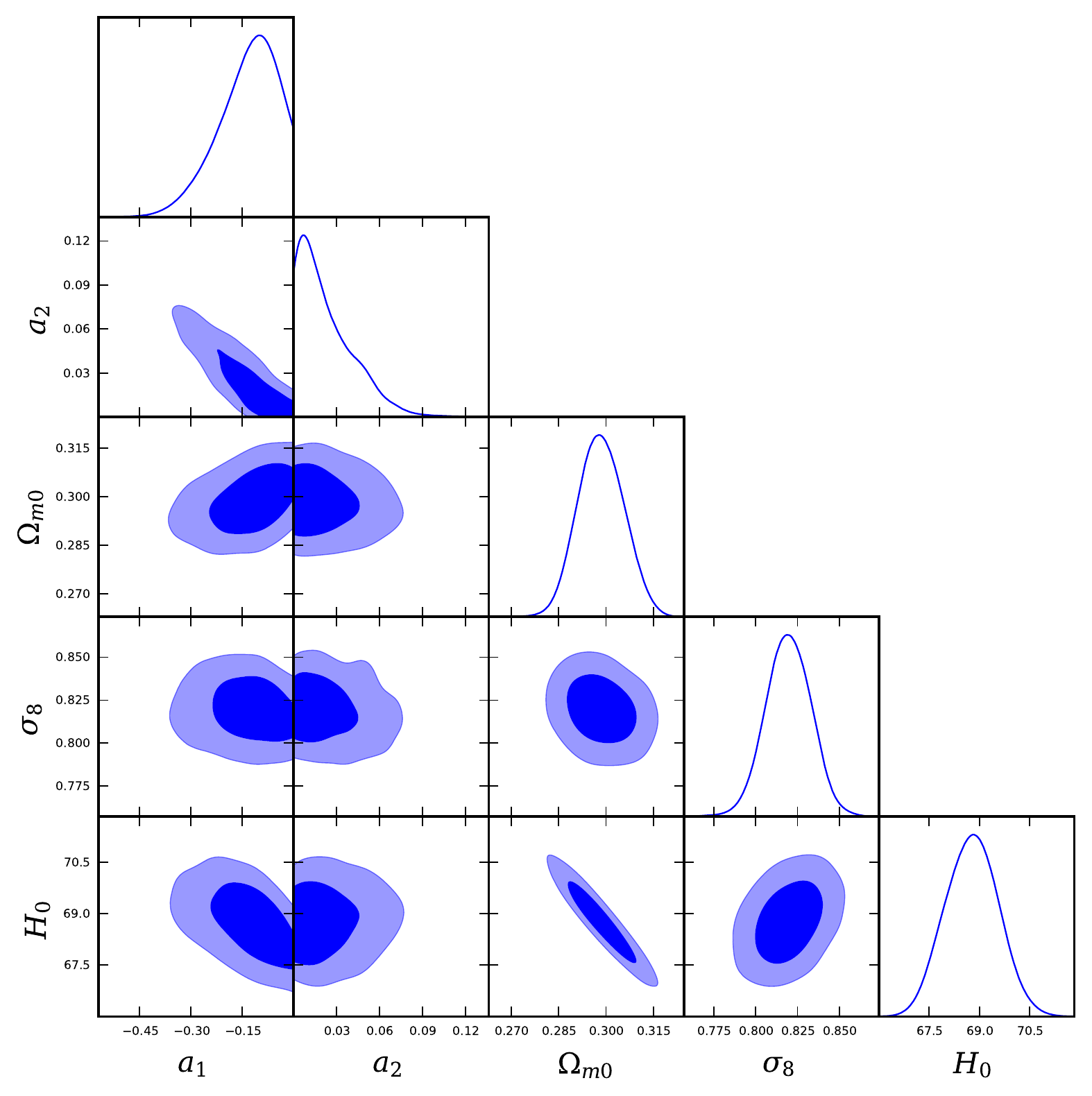}
\caption{\textit{The figure displays the 68.3\% and 95.4\% confidence-level
contour plots for different combinations of the model parameters of the
Nesseris-Perivolaropoulos parametrization (\ref{nnl}) using the combined analysis Planck
TT, TE, EE $+$ lowTEB $+$ JLA $+$ BAO $+$ RSD $+$ WL $+$ CC $+$ $H_{0}$. }}%
\label{fig:NP}%
\end{figure}

\begingroup
\begin{table}
\begin{tabular}
[c]{ccccc}\hline\hline
Parameters &  Priors &Mean with errors & Best fit\\\hline
$\Omega_{c} h^{2}$ & $[0.01, 0.99]$ & $0.1182_{- 0.0012- 0.0023}^{+
0.0011+ 0.0023}$ & $0.1186$\\
$\Omega_{b} h^{2}$ & $[0.005, 0.1]$ & $0.0223_{- 0.0001- 0.0003}^{+
0.0001+ 0.0003}$ & $0.0223$\\
$100\theta_{MC}$ & $[0.5, 10]$ &$1.0406_{- 0.0003- 0.0006}^{+
0.0003+ 0.0006}$ & $1.0405$\\
$\tau$ & $[0.01, 0.8]$ & $0.0677_{- 0.0179- 0.0317}^{+ 0.0161+
0.0339}$ & $0.0616$\\
$n_{s}$ & $[0.5, 1.5]$ &$0.9762_{- 0.0043- 0.0077}^{+ 0.0039+
0.0080}$ & $0.9732$\\
$\mathrm{{ln}}(10^{10} A_{s})$ & $[2.4, 4]$ & $3.0741_{- 0.0344- 0.0624}^{+ 0.0309+ 0.0677}$ & $3.0669$\\\hline
$w_{0}$ & $[-2, 0]$ & $-1.0533_{- 0.0321- 0.0668}^{+ 0.0362+
0.0639}$ & $-1.0376$\\
$w_{b}$ & $[0, 2]$ &$0.0971_{- 0.0971- 0.0971}^{+ 0.0160+
0.1694}$ & $0.0201$\\\hline
$\Omega_{m0}$ & $-$ &$0.2987_{- 0.0070- 0.0143}^{+
0.0071+ 0.0145}$ & $0.3011$\\
$\sigma_{8}$ & $-$ & $0.8212_{- 0.0147- 0.0243}^{+
0.0125+ 0.0278}$ & $0.8189$\\
$H_{0}$ & $-$ & $68.7539_{- 0.8010- 1.5505}^{+ 0.7703+
1.6054}$ & $68.5523$\\\hline
$\chi^{2}_{min}$ & $-$ & $-$ & 13722.95\\\hline\hline
\end{tabular}
\caption{The table summarizes the observational constraints of the free and
derived parameters of the Model D (The new parametrization) for the combined
observational data Planck TT, TE, EE $+$ lowTEB $+$ JLA $+$ BAO $+$ RSD $+$ WL
$+$ CC $+$ $H_{0}$. We note that, $\Omega_{m0}=\Omega_{c0}+\Omega_{b0}$.}%
\label{tab:YPP}%
\end{table}
\endgroup

\begin{figure}
\includegraphics[width=8cm,height=7.5cm]{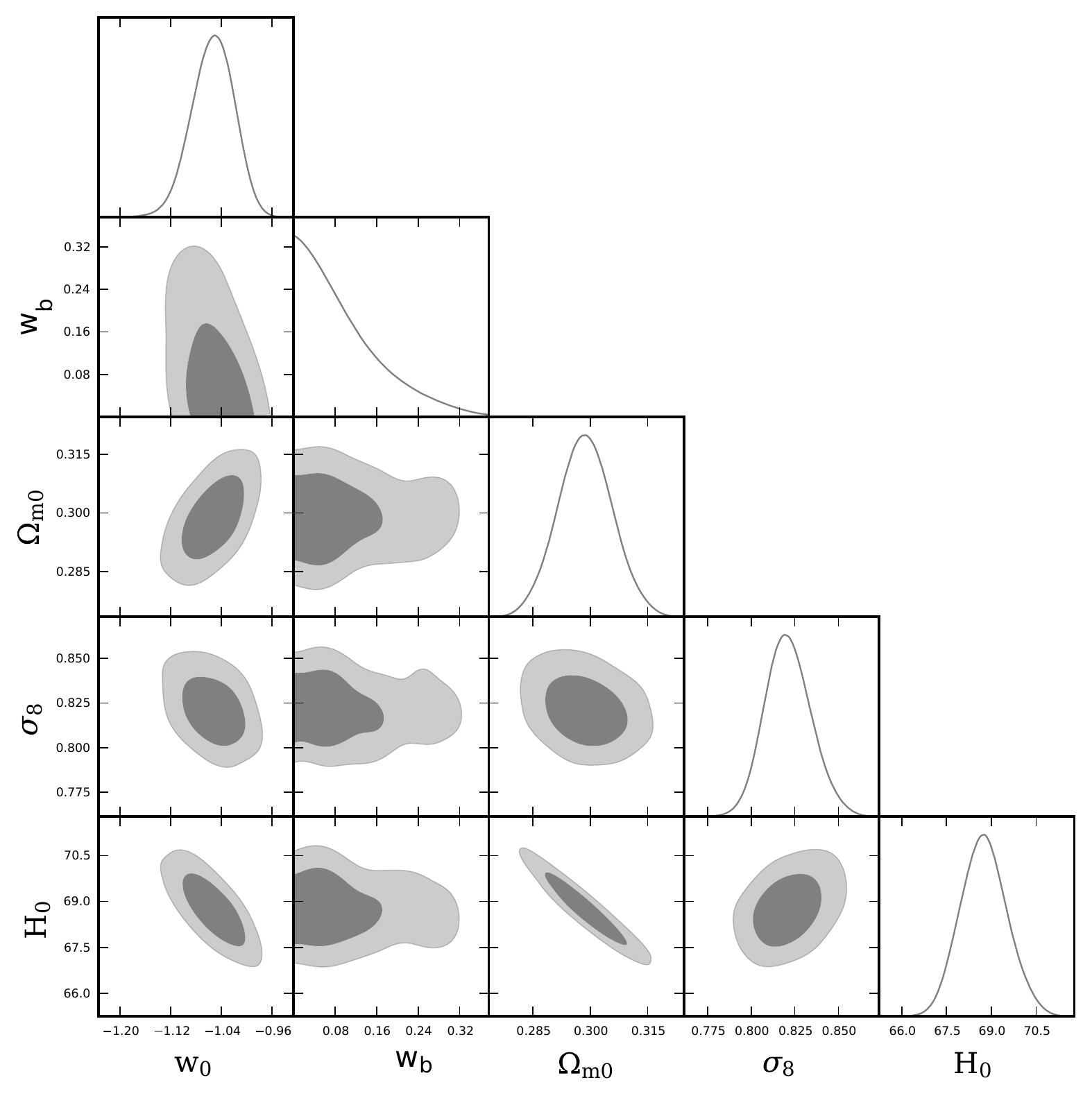}
\caption{\textit{The figure displays the 68.3\% and 95.4\% confidence-level
contour plots for different combinations of the model parameters of the
new parametrization (\ref{model-d}) using the combined analysis Planck
TT, TE, EE $+$ lowTEB $+$ JLA $+$ BAO $+$ RSD $+$ WL $+$ CC $+$ $H_{0}$. }}%
\label{fig:YPP}%
\end{figure}

\begin{figure}
\includegraphics[width=7.5cm,height=7.0cm]{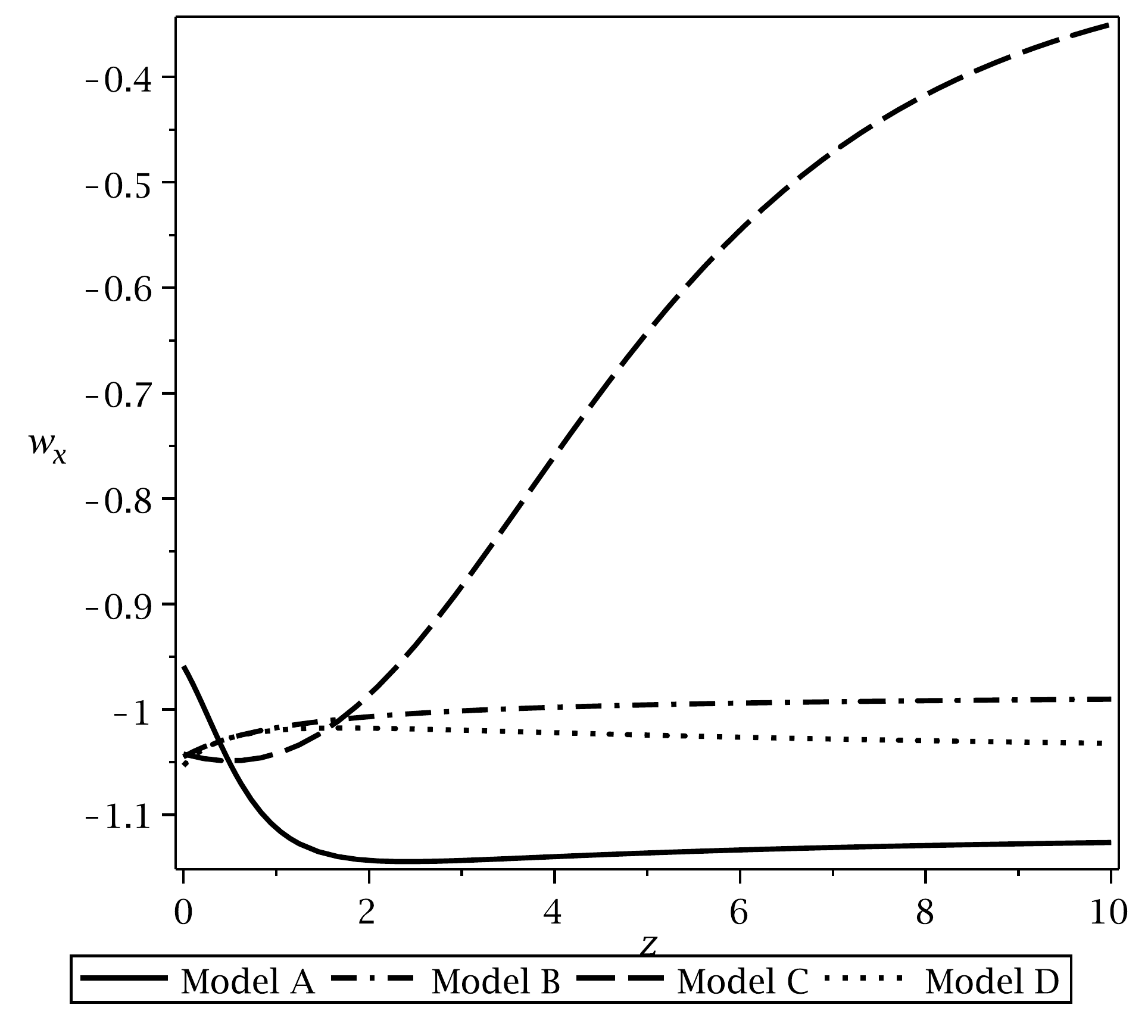} \caption{\textit{The
evolution of the dark energy equation of state for models $A-D$ using the mean
values from Planck TT, TE, EE $+$ lowTEB $+$ JLA $+$ BAO $+$ RSD $+$ WL $+$ CC
$+$ $H_{0}$. }Solid line is for model A, dash-dot line for model B, dash-dash
line for model C and dot-dot line is for model D. From the plot it can be seen
that all the models cross the phantom divine line in the past while only model
A provides an equation of state parameter with value bigger than minus one at
the present time. }%
\label{fig:eos}%
\end{figure}

\section{Observational data sets}

\label{sec-data}

To constrain the four parametric dark energy models we have used the 
latest astronomical data from several sources. In the following 
we briefly describe the employed data sets and their sources. 

\begin{itemize}
\item The \textit{Cosmic Microwave Background data (CMB)} given by the Planck
2015 collaboration \cite{ref:Planck2015-1, ref:Planck2015-2}~\ have been used
here. Here we combine the likelihoods $C_{l}^{TT}$, $C_{l}^{EE}$, $C_{l}^{TE}$
in addition to the low$-l$ polarization data. We identify this 
data with the notation $C_{l}^{TE}+C_{l}^{EE}+C_{l}^{BB}$.

\item The joint light curve (JLA) sample \cite{Betoule:2014frx} provides with 740
\textit{Supernovae Type Ia (SNIa)} data in the redshift range
$z\in\lbrack 0.01,1.30]$.

\item The \textit{Baryon acoustic oscillations (BAO) distance measurements~}%
from various collaborations are used. In particular\textit{,} for the BAO data
we consider the estimated ratio of $r_{s}/D_{V}$ as a `standard ruler' in which
$r_{s}$ is the comoving sound horizon at the baryon drag epoch and $D_{V}$ is
the effective distance determined by the angular diameter distance $D_{A}$ and
Hubble parameter $H$ taking the form $D_{V}(z)=\left[  (1+z)^{2}D_{A}(a)^{2}\frac{z}%
{H(z)}\right]  ^{1/3}$. For this data set, we use four BAO points, namely the
6dF Galaxy Survey (6dFGS) measurement at $z_{\emph{\emph{eff}}}=0.106$
\cite{ref:BAO1-Beutler2011}, the Main Galaxy Sample of Data Release 7 of Sloan
Digital Sky Survey (SDSS-MGS) at $z_{\emph{\emph{eff}}}=0.15$
\cite{ref:BAO2-Ross2015}, and the CMASS and LOWZ samples from the latest Data
Release 12 (DR12) of the Baryon Oscillation Spectroscopic Survey (BOSS) at
$z_{\mathrm{eff}}=0.57$ \cite{ref:BAO3-Gil-Marn2015} and $z_{\mathrm{eff}%
}=0.32$ \cite{ref:BAO3-Gil-Marn2015}.

\item We use the \textit{Redshift space distortion (RSD)} data from two
observational surveys which include the CMASS sample with an effective
redshift of $z_{\mathrm{eff}}=0.57$ \cite{ref:BAO3-Gil-Marn2015} 
and the LOWZ sample with an effective
redshift of $z_{\mathrm{eff}}=0.32$~\cite{ref:BAO3-Gil-Marn2015}.

\item Moreover, we consider the\textit{ weak gravitational lensing} (WL) data
from blue galaxy sample compliled from Canada$-$France$-$Hawaii Telescope
Lensing Survey (CFHTLenS) \cite{Heymans:2013fya,Asgari:2016xuw}.

\item Finally, \textit{cosmic chronometers (CC) plus the local value of the
Hubble parameter (CC $+$ $H_{0}$)\thinspace\ }are added with the previous data
sets. The cosmic chronometers data compile $30$ measurements of the Hubble
parameter values in the redshift interval $0<z<2$ \cite{Moresco:2016mzx} and
they are determined through a model independent manner, hence, they are
believed to provide with crucial information 
from the cosmological model. The local
value of the Hubble parameter from the Hubble Space Telescope yields
$H_{0}=73.02\pm1.79$ km/s/Mpc with 2.4\% precision as determined recently in
\cite{Riess:2016jrr}.
\end{itemize}

\section{Results of the observational constraints}

\label{sec-results}

Let us summarize the main observational constraints extracted from the
parametric dark energy models using the combined astronomical data
\textit{Planck TT, TE, EE $+$ lowTEB $+$ JLA $+$ BAO $+$ RSD $+$ WL $+$ CC $+$
$H_{0}$}. The analysis is based on the total likelihood $\mathcal{L}\propto
e^{-\chi^{2}_{tot}/2}$, where $\chi^{2}_{tot} = \sum_{j} \chi^{2}_{j}$, and
$j$ runs over all the data sets such as CMB (\textit{Planck TT, TE, EE $+$ lowTEB}), \textit{JLA, BAO,
RSD, WL, CC} and $H_{0}$. For the statistical analysis, we use the Markov Chain Monte Carlo package \texttt{cosmomc}, a Metropolis-Hastings algorithm \cite{Lewis:2002ah, Lewis:2013hha} which extracts the cosmological constraints in an efficient way following a convergence criteria of the model parameters based on the Gelman-Rubin statistics \cite{Gelman-Rubin}. Precisely, for each dark energy model, we modify a fast Boltzmann code known as CAMB (Cosmic Anisotropy Microwave Background) within the \texttt{cosmomc} pacakge to solve the evolution equations and consequently, the $\chi^{2}_{tot}$ function for the combined data set. Finally, calling \texttt{cosmomc} we explore the parameters space for the models.  In all
parametric models, we have the following eight-dimensional parameters space
\begin{align}
\mathcal{P} \equiv\Bigl\{\Omega_{b}h^{2}, \Omega_{c}h^{2}, 100 \theta_{MC}, \tau,
P_{1}, P_{2}, n_{s}, log[10^{10}A_{s}]\Bigr\}, \label{eq:parameter_space}%
\end{align}
where $P_{1}$, $P_{2}$ are the model parameters and the remaining parameters
($\Omega_{b}h^{2}$, $\Omega_{c}h^{2}$, $100 \theta_{MC}$, $\tau$, $n_{s}$, $A_{s}$)
are respectively the baryons density, cold dark matter density, ratio of sound
horizon to the angular diameter distance, optical depth, scalar spectral
index, and the amplitude of the initial power spectrum. In the following we
describe the results for each model.

\begin{itemize}
\item \textbf{Model A: }We constrain this parametrization using the combined
observational data mentioned above. The Table \ref{tab:BA} summarizes the
observational constraints of the free parameters ($w_{0}$, $w_{a}$) of this
model as well as the derived parameters ($\sigma_{8}$, $\Omega_{m0}$, $H_{0}$)
and the Figure \ref{fig:BA} displays the 68.3\% ($1\sigma$) and 95.4\%
($2\sigma$) confidence-level contour plots for several combinations of the
free and derived parameters. Our analysis shows that the best fit and the mean
value of the dark energy equation of state, $w_{x}$, for this parametrization
show its quintessential nature, however, one cannot exclude the possibility
of the dark energy equation of state to be phantom. In fact at 1$\sigma$
lower confidence-level, $w_{x}<-1$, but that is very close to the cosmological constant
$w_{x}=-1$. However, 
we note that the observed tension between the 
different values of $H_{0}$ reported from the
local ($H_{0}=67.27\pm0.66$ Km s$^{-1}$ Mpc$^{-1}$) \cite{Riess:2016jrr} and
the global measurements ($H_{0}=73.24\pm1.74$ Km s$^{-1}$ Mpc$^{-1}$)
\cite{Ade:2015xua} may be reconciled for this model. 
This might be considered to be one of
the interesting properties in the dynamical DE models. We finally 
note that the estimation of $\sigma_8$ 
 ($= 0.8197_{- 0.0132}^{+ 0.0144}$ at 
68.3\% confidence-level) for this model is close to 
the Planck's $\Lambda$CDM based estimation, in particular when lensing is added to either Planck TT+lowP or 
Planck TT, TE, EE+lowP. Precisely the estimated values of $\sigma_8$ are $\sigma_8= 0.8149 \pm 0.0093$ at 68.3\% confidence-level (Planck TT+lowP+lensing) \cite{Ade:2015xua} and $\sigma_8= 0.8150 \pm 0.0087$ at 68.3\% confidence-level (Planck TT, TE, EE+lowP+lensing) \cite{Ade:2015xua}. The addition of external data (BAO+JLA+$H_0$) to both Planck TT+lowP+lensing or Planck TT, TE, EE+lowP+lensing does not offer any significant changes to $\sigma_8$, see \cite{Ade:2015xua}. 

\item \textbf{Model\ B:} We summarize the observational constraints for this
model in Table \ref{tab:ZhangI} and in Figure \ref{fig:ZhangI} we show the
corresponding contours at 68.3\% and 95.4\% confidence-levels. For the
analysis we have used the same observational data as mentioned. Here we see
that both best fit and the mean values of the dark energy equation of state
$w_{x}$, are phantom! In fact, within 1$\sigma$ confidence-level, the dark
energy equation of state remains phantom, i.e. $w_{x}<-1$, 
however, within $2\sigma$ confidence-level it is
quintessential. One can further notice that the model can be considered to
reconcile the current tension on $H_{0}$ within a confidence level slightly greater
than $1\sigma$. Further, we also see that the value of $\sigma_8$ for this model
is very similar to the estimated value of $\sigma_8$ for Model $A$ and also very close to the Planck's $\Lambda$CDM based estimation in presence of lensing \cite{Ade:2015xua}. 

\item \textbf{Model\ C: }As far as the authors are concerned with
\cite{Nesseris:2005ur}, this model has not been constrained yet due to its
very complicated nature. Using the monte-carlo simulation and the
same observational data, we found that the model can be constrained. Thus, we
present the observational constraints of this model summarized in Table
\ref{tab:NP} while Figure \ref{fig:NP} shows the $1\sigma$ and $2\sigma$
confidence-level contour plots for several combinations of the free and
derived parameters. Using the constrained values of the free parameters
$a_{1}$ and $a_{2}$ and the derived parameter $\Omega_{m0}$, the current value
of the dark energy equation of state, $w_{0}$, can be reproduced. The mean and
the best fit values of $w_{0}$ up to 4 places of decimal can be found to be
$-1.0426$, and $-1.0401$, respectively, which show that both mean and the best
fit values of the dark energy equation of state for this model are very close
to the cosmological constant boundary `$-1$'. The current tension on $H_{0}$
can also be relieved in this dynamical model in the confidence level greater
than $1\sigma$. Concerning the estimated value of $\sigma_8$, this model also follows similar trend to that of Model $A$ and $B$ and hence Planck's estimation in presence of lensing \cite{Ade:2015xua}. 

\item \textbf{Model\ D: } This is a new model proposed in this work. Using the
same combined analysis applied to the previous models we constrain this model
with results summarized in Table \ref{tab:YPP} and the corresponding contour
plots at $1\sigma$ and $2\sigma$ confidence-levels displayed in Figure
\ref{fig:YPP}. We see that the current value of the dark energy equation of
state, $w_{0}$, for this model crosses the phantom divide line, moreover,
within $1\sigma$ confidence-level, $w_{0}<-1$. However, the results on $w_{0}$
show that within $2\sigma$ confidence-level, the behaviour of $w_{0}$ becomes
quintessential. Concerning the current tension on $H_{0}$, this model follows
a similar trend to Models $B$ and $C$. However, one can see that the 
model allows a slight greater value of $\sigma_8$ in compared to its
value obtained for the Models $A-C$ as well as from the $\Lambda$-cosmology 
\cite{Ade:2015xua}. One can see that Model $D$ is slightly different from the other 
three models, but however, the difference is not much significant. 
\end{itemize}

We close this section with Figure \ref{fig:eos} that shows the behaviour of
$w_{x} (z)$ for models $A-D$ using the mean values of the free and derived
parameters from the said analysis. The figure shows that at late time, the
models $B-D$ cross the cosmological constant boundary leading to a phantom
dark energy dominated universe while for model $A$ the dark energy equation of
state still lies just above the cosmological constant line, which hints for a
quintessence dark energy dominated universe.

\begin{figure}
\includegraphics[width=8cm,height=7.5cm]{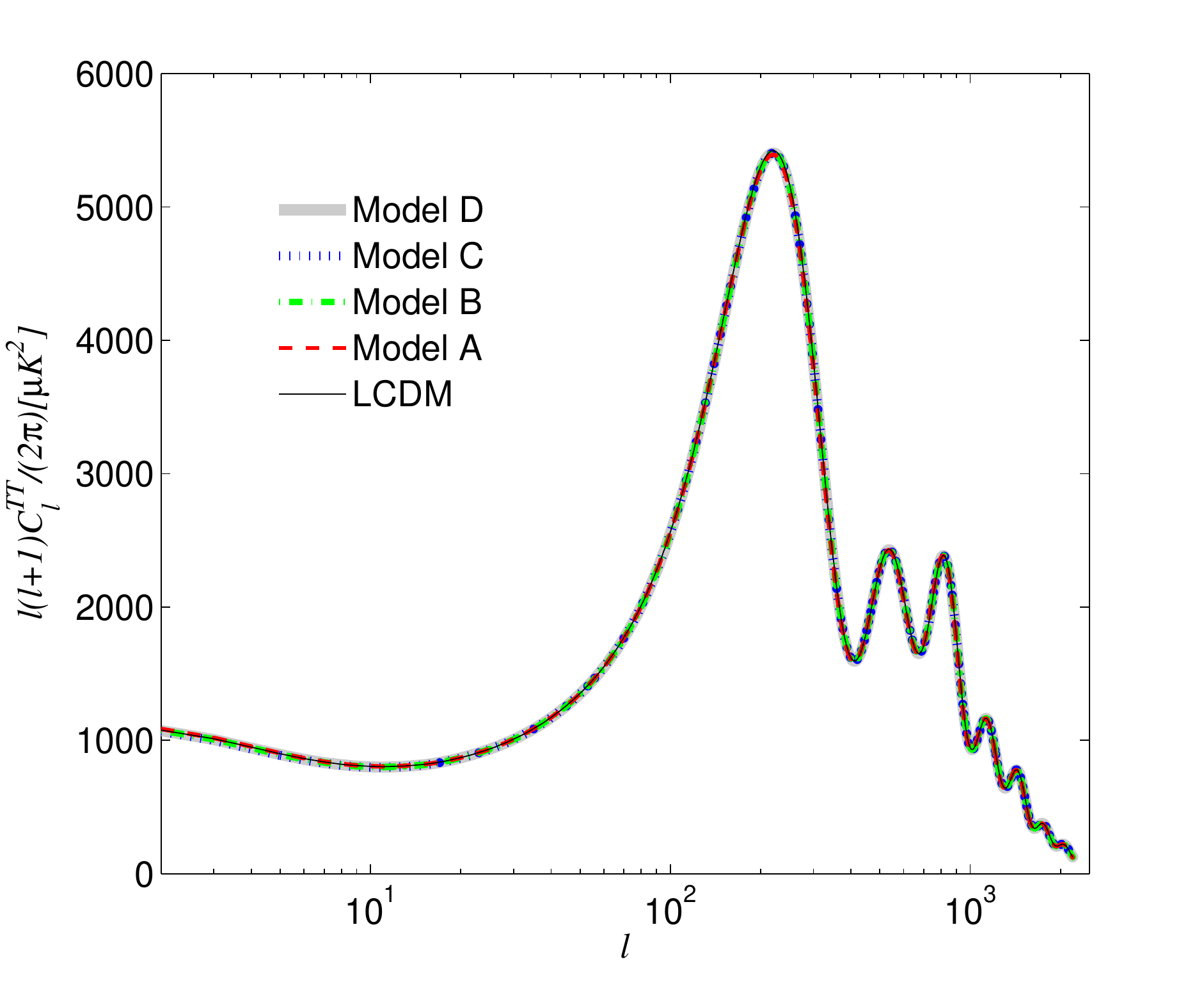}
\caption{\textit{Angular CMB temperature power spectra for the $\Lambda
$-cosmology (LCDM) and the parametric dark energy models $A-D$ have been displayed
using the mean values of the model parameters from Planck TT, TE, EE $+$
lowTEB $+$ JLA $+$ BAO $+$ RSD $+$ WL $+$ CC $+$ $H_{0}$. We see that the
curves in the plot are almost indistinguishable from each other meaning their
closeness to each other.}}%
\label{cmbpower}%
\end{figure}\begin{figure}
\includegraphics[width=8cm,height=7.5cm]{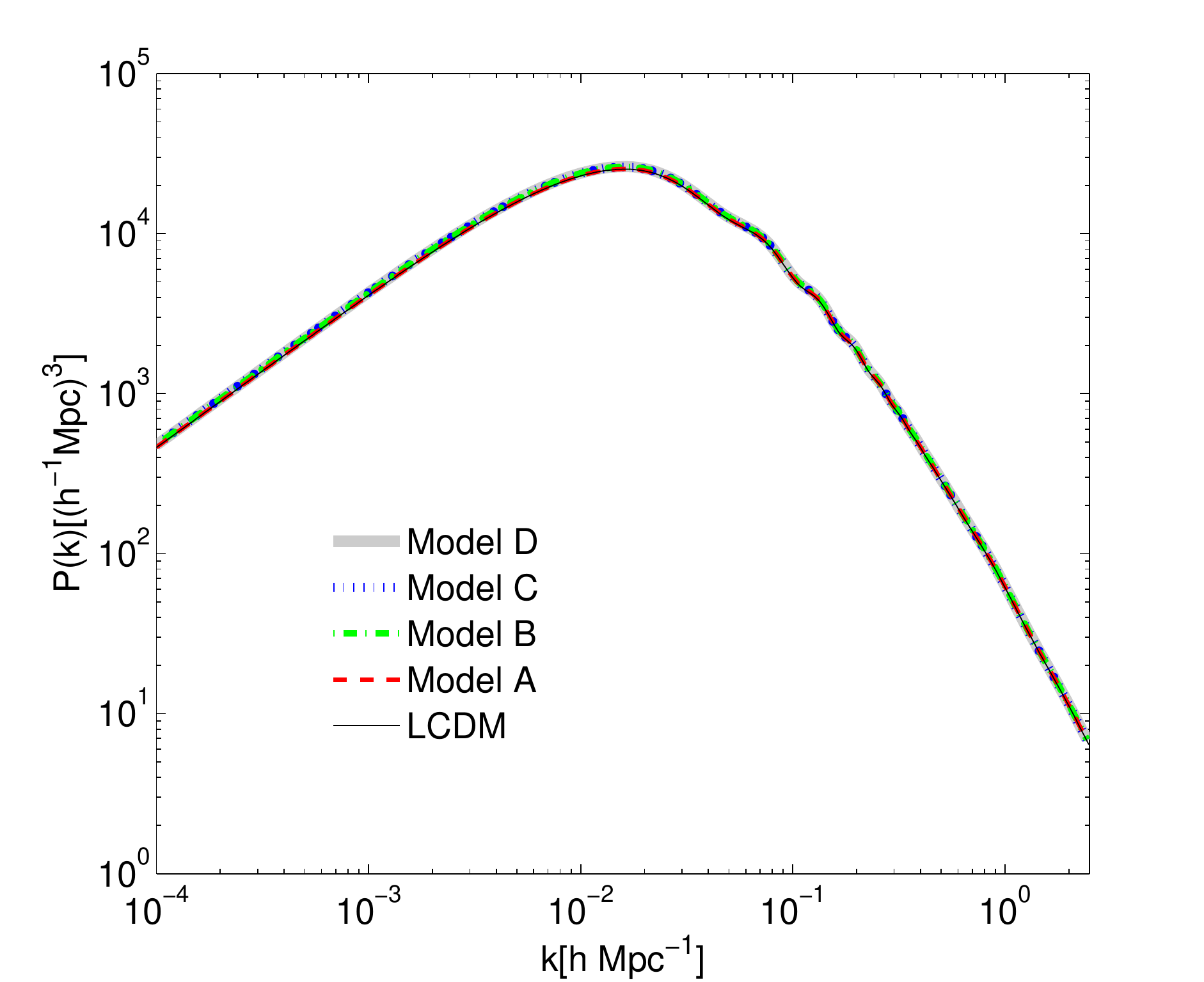}
\caption{\textit{Matter
power spectra for the $\Lambda$-cosmology (LCDM) and the parametric dark energy
models $A-D$ using the mean values of the model parameters from Planck TT, TE,
EE $+$ lowTEB $+$ JLA $+$ BAO $+$ RSD $+$ WL $+$ CC $+$ $H_{0}$. The curves in
the plot cannot be distinguished from each other which means that the models
at large scales are very close to each other.} }%
\label{mpower}%
\end{figure}

\begin{figure}
\includegraphics[width=8cm,height=7.0cm]{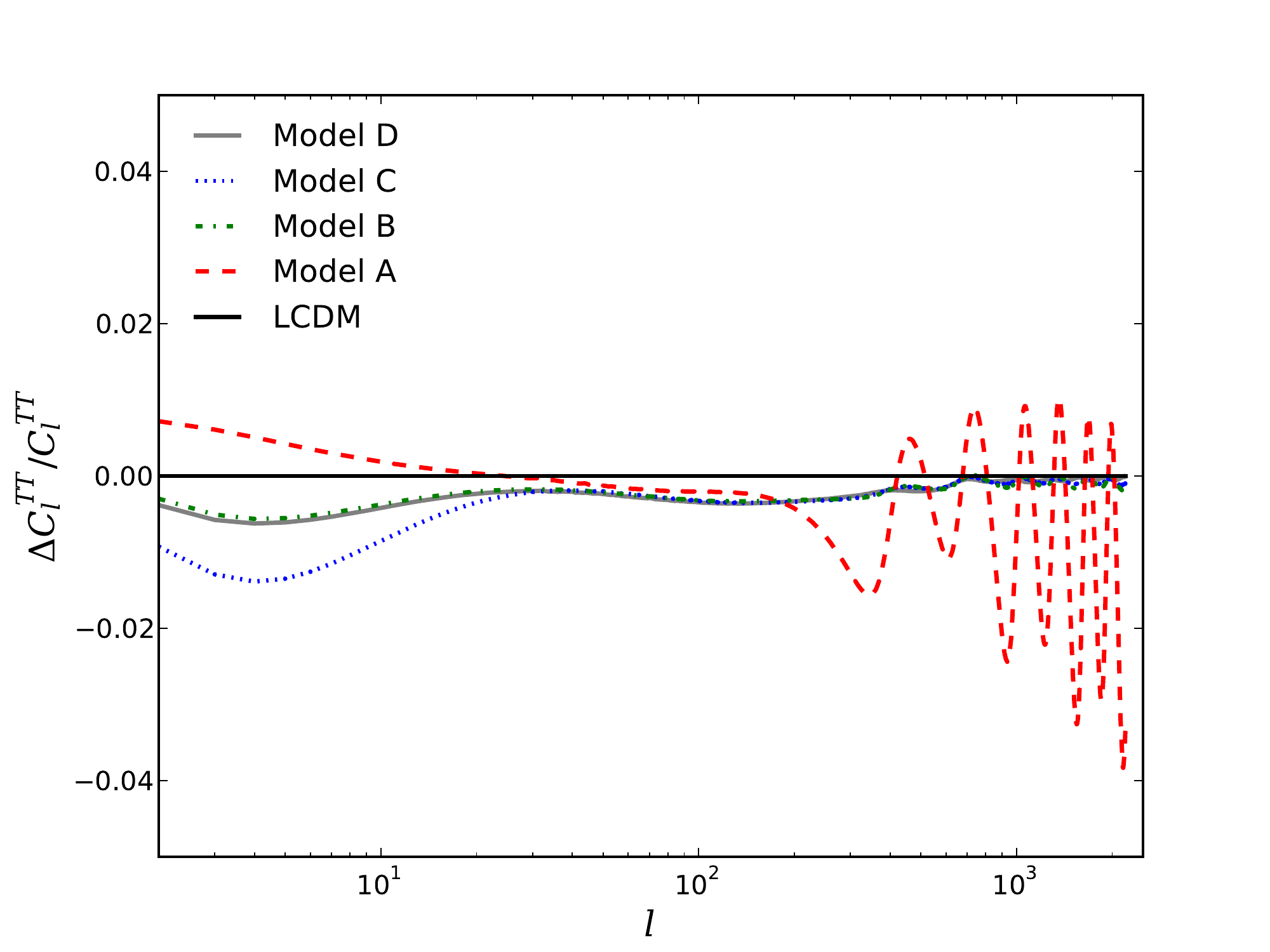}
\includegraphics[width=8cm,height=7.0cm]{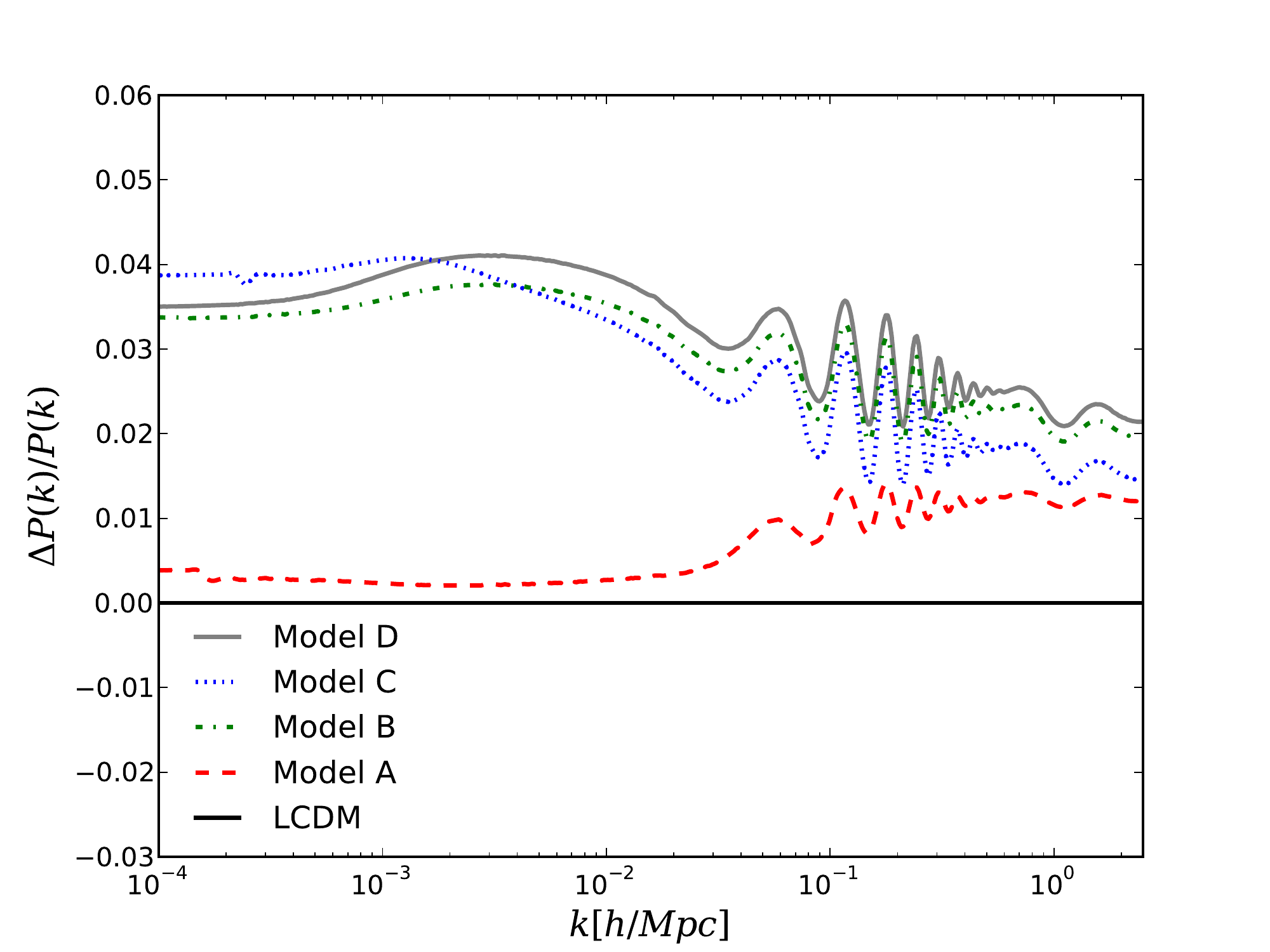}
\caption{\textit{The figure displays the relative deviations in the CMB TT spectra ($\Delta C_l^{TT}/C_l^{TT}$) and the matter power spectra ($\Delta P_k^{TT}/P_k^{TT}$) from the flat $\Lambda$-cosmology (LCDM) using the combined analysis Planck TT, TE, EE $+$ lowTEB $+$ JLA $+$ BAO $+$ RSD $+$ WL $+$ CC $+$ $H_{0}$. Here, $\Delta C_l^{TT} =C_{l}^{TT}\bigl|_{model} \, -\, C_{l}^{TT}\bigl|_{LCDM}$ and $C_l^{TT} = C_{l}^{TT}\bigl|_{LCDM}$. Similarly, $\Delta P(k) =P(k)\bigl|_{model}- P(k)\bigl|_{LCDM}$ and $P(k) = P(k)|_{LCDM}$. One can notice that the parametric dark energy models in (\ref{BA}), (\ref{Zhang1}), (\ref{nnl}) and (\ref{model-d}) have deviations from the flat $\Lambda$-cosmology, however, from the deviations as seen in both the plots one can conclude that the models are not far from the $\Lambda$-cosmology. }}
\label{fig:ratio}
\end{figure}

\section{Summary and conclusions}

\label{discuss}

The theory of dark energy has been one of the central themes for describing
the current accelerating phase of the universe. From the simplest cosmological
constant to several complicated dark energy models are included in this
category. Parametric dark energy models are one of them where the evolution of
the dark energy models are described through their equations of state. In this work we
consider some parametric dark energy models, namely, Model $A$, Model $B$, Model $C$
and Model $D$ in order to understand the late-evolution of the universe. Among
these models although Model $A$ has been tested earlier with the observational
data while the others are not, but still we tested Model $A$ again in order to
update its observational constraints from the latest observational data that
we employ in this work. To provide a better constarints on the models, we
consider the large scale behaviour of the models and examine the models with
the combined analysis of cosmic microwave background radiation, Supernove Type
Ia, baryon acoustic oscillations, redshift space distortion data, weak
gravitational lensing data, cosmic chronometers plus the local Hubble constant
measured at 2.4\% precision. Finally, we use the Markov Chain Monte Carlo package \texttt{cosmomc}, a Metropolis-Hastings algorithm \cite{Lewis:2002ah, Lewis:2013hha} to fit all the models.

Our analysis shows that at present epoch the Models $B$, $C$ and $D$ may allow the
dark energy equation of state to cross the cosmological constant boundary
`$w_{x}=-1$', however, the equation of state for those models always stays
very close to `$-1$'. For Model $A$ the dark energy equation of state at present
is quintessential. Moreover, we find that within 68.3\% confidence region,
Model $A$ can alleviate the current tension on $H_{0}$ as suggested from the
local \cite{Riess:2016jrr} and global measurements \cite{Ade:2015xua} while
the other models allow such property in a greater confidence level than
68.3\%. We have also found that the values of $\sigma_8$ for all models are almost similar and they are in agreement with the $\Lambda$CDM based 
Planck's estimation \cite{Ade:2015xua}, although for model $D$, a slightly higher value of $\sigma_8$ is preferred which however is not a serious issue to make a strong comment.  In summary we find that the observational constraints on the models
suggest that the current value of the dark energy equation of state is close
to $w_{x}=-1$, but effectively the models differ from that of $\Lambda
$-cosmology for non-null values of $w_{a}$, $w_{2}$, $w_{b}$, $a_{1}$ and
$a_{2}$.

Furthermore, using the same observational data described above, we derive
that $\chi_{min}^{2}=13723.806$ ($\Lambda$CDM), $\chi_{min}^{2}=13720.556$
(Model $A$), $\chi_{min}^{2}=13722.266$ (Model $B$), $\chi_{min}^{2}=13723.394$
(Model $C$) and $\chi_{min}^{2}=13722.95$ (Model $D$). We observe that all the
models fit the data in a similar way. It is important to mention that the
parametric dark energy models have more degrees of freedom in compared to
the $\Lambda$-cosmology. However, by using 
the Akaike information criterion
\cite{Akaike1974, sugu} to study the relation between the models with different
degrees of freedom we can say that those models are statistically equivalent,
at least with the current observational data.

Last but not least, in Figs. \ref{cmbpower} and \ref{mpower} we respectively
show the angular CMB tempertaure power spectra and the matter power spectra
for the parametric dark energy models in compared to the flat $\Lambda
$-cosmology from which we observe that the parametric models are very close to
each other as well as also with the $\Lambda$-cosmology. However, since the models are dynamical and different from one another, it is anticipated that they should exhibit deviation from each other (whatever small it could be) as well as from the $\Lambda$-cosmology. We note that the differences of the models are not detected from Fig. \ref{cmbpower} and Fig. \ref{mpower} while from the Fig. \ref{fig:ratio} measuring the relative deviations of the models to that of the $\Lambda$-cosmology in terms of the CMB TT spectra and  matter power spectra, one can see that the models have deviations from each other as well as from $\Lambda$-cosmology. 
That is an expected
result because in the past those parametric models should almost have a constant value for the equation of state parameter.

\section*{Acknowledgements}
The authors thank the referee for some essential comments
to improve the work.
The work of WY has been supported by the National Natural Science Foundation of China under Grants No. 11705079 and No. 11647153. SP
acknowledges the partial support from SERB-NPDF (File No. PDF/2015/000640). AP
acknowledges the financial support \ of FONDECYT grant no. 3160121.  The authors 
acknowledge the use of publicly available Monte Carlo Markov Chain package \texttt{cosmomc}.

\bigskip

\end{document}